\documentclass[sn-mathphys,Numbered,pdflatex]{sn-jnl}%

\usepackage{graphicx}%
\usepackage{multirow}%
\usepackage{amsmath,amssymb,amsfonts}%
\usepackage{amsthm}%
\usepackage{mathrsfs}%
\usepackage[title]{appendix}%
\usepackage{xcolor}%
\usepackage{textcomp}%
\usepackage{manyfoot}%
\usepackage{booktabs}%
\hypersetup{hypertexnames=false}

\usepackage{siunitx}
\usepackage{gensymb}
\usepackage[version=4]{mhchem}

\usepackage{cleveref}
\usepackage{subfig}
\crefname{figure}{Fig.}{Figs.}
\crefname{equation}{Eq.}{Eqs.}

\let\origcite\cite
\def\cite#1{\unskip~\origcite{#1}}
\let\origcitep\citep
\def\citep#1{\unskip~\origcitep{#1}}
\begin{document}

\title{Snakes in the Plane: Controllable Gliders in a Nanomagnetic Metamaterial}
\author*[1]{\fnm{Arthur} \sur{Penty}}
 \email{arthur.penty@ntnu.no}
\author[1]{\fnm{Johannes H.} \sur{Jensen}}
\author[2]{\fnm{Ida} \sur{Breivik}}
\author[2]{\fnm{Anders} \sur{Str\o mberg}}
\author[2]{\fnm{Erik} \sur{Folven}}
\author[1]{\fnm{Gunnar} \sur{Tufte}}
\affil[1]{\orgdiv{Department of Computer Science}, \orgname{Norwegian University of Science and Technology}, Trondheim, Norway}

\affil[2]{\orgdiv{Department of Electronic Systems}, \orgname{Norwegian University of Science and Technology}}

\abstract{ %
The magnetic metamaterials known as Artificial Spin Ice (ASI) are promising candidates for neuromorphic computing, composed of vast numbers of interacting nanomagnets arranged in the plane.
Every computing device requires the ability to transform, transmit and store information. 
While ASI excel at data transformation, reliable transmission and storage has proven difficult to achieve.
Here, we take inspiration from the Cellular Automaton (CA), an abstract computing model reminiscent of ASI.
In CAs, information transmission and storage can be realised by the ``glider'', a simple structure capable of propagating while maintaining its form.
Employing an evolutionary algorithm, we search for gliders in pinwheel ASI and present the simplest glider discovered: the ``snake''. Driven by a global field protocol, the snake moves strictly in one direction, determined by its orientation. 
We demonstrate the snake, both in simulation and experimentally, and analyse the mechanism behind its motion.
The snake provides a means of manipulating a magnetic texture in an ASI with resolution on the order of \qty{100}{\nm}, which could in turn be utilised to precisely control other magnetic phenomena.
The integration of data transmission, storage and modification into the same magnetic substrate unlocks the potential for ultra-low power computing devices.

}

\maketitle

\section*{Introduction}
Artificial Spin Ice (ASI) are metamaterials composed of many nanomagnets arranged on a 2D lattice.
The nanomagnets behave as binary artificial spins, which interact through dipolar coupling.
By altering the placement and orientation of the nanomagnets, a wide variety of emergent behaviour can be achieved, e.g., long-range ordering\citep{Macedo2018,barrows2019}, magnetic monopoles and Dirac strings\citep{Mengotti2011}, charge screening\citep{farhan2016}, and adherence to the ice rules\citep{Wang2006,gilbert2014}.
Recently, a new class of field protocols called ``astroid clocking'' has been developed that selectively switch nanomagnets within the ensemble\citep{jensen2024clocked}, offering greater control of the time-evolution (dynamics) of the ASI.

A potential application for ASI is as a substrate for neuromorphic computing\citep{Jensen2018}, specifically within a reservoir computing framework\citep{jensen2020,Hon_ASIrc2021,gartside2022reconfigurable,Stenning2024}.
Despite promising results\citep{jensen2020,gartside2022reconfigurable}, a major challenge has been a lack of short-term memory\citep{Hon_ASIrc2021,Stenning2024}, which either limits their utility or necessitates external support circuitry, such as delay line memories.
External memories add significant overhead and reduce efficiency, since the reservoir state must be constantly read out and stored.
There appears to be a fundamental computational property which has thus far eluded ASI computing approaches.

Conceptually, ASI are systems of a large number of simple elements governed by local interactions.
This property is a defining characteristic of the Cellular Automaton (CA), a simple model of distributed computation from which complex behaviour emerges.
CAs have been the subject of intense study since the early days of computing as a means to investigate fundamental principles of computation\citep{von66BURKS,Codd1968,Banks1971}.
A wide range of CAs have been proven to be computationally universal, from simple elementary 1D CAs\citep{Cook2004}, to more complex 2D CAs\citep{Codd1968,Banks1971,Rendell2016}.
Here, we consider ASI through the lens of CA theory, which provides key insights into the computational properties of ASI.

Fundamentally, any computing substrate must support the transmission, storage and modification of information\citep{LANGTON91}.
In one of the most famous 2D CAs, Conway's Game of Life\cite{Gardner1970}, information transmission can be realised by the \emph{glider}: a small 5-cell structure that can travel indefinitely through the CA unless it collides with another structure.
Gliders and other glider-like structures can be used to transmit signals and form the basis of a CA computing model\citep{morita2022gliders}.
Additionally, the reliable movement of the glider can be used as a basis for memory, since information can flow through the system without degrading.
The concept of a glider can be generalised beyond the Game of Life, to any structure that can move through a substrate while maintaining its form, i.e., translation.

Can gliders be found in an ASI?
Here, we show that the answer is a definite yes.
Using an Evolutionary Algorithm (EA), we search for gliders in pinwheel ASI and present the simplest glider structure discovered, which we term the ``snake''.
The snake glider can move either left or right, with the direction decided by the orientation of its structure.
We verify the snake both in simulation and in experiment, and analyse the mechanism behind its movement.
The snake is surprisingly robust, enabling reliable information transfer and storage within the ASI metamaterial.

\section*{Results}
\begin{figure*}
    \centering
    \includegraphics[width=\textwidth]{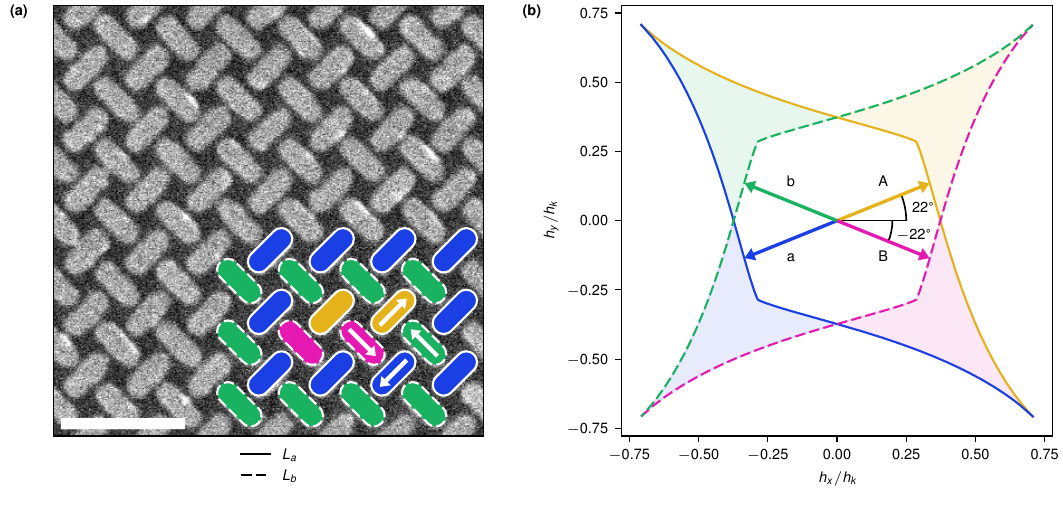}
    \caption{Astroid clocking of pinwheel ASI.
    (a) A SEM image of a pinwheel ASI.
    The inset in the lower-right corner illustrates magnetisation with colours corresponding to magnetisation directions, as indicated by the white arrows.
    The inset shows a small rightwards (orange-pink) domain surrounded by a leftwards (blue-green) background domain.
    Scale bar: \qty{500}{\nm}
    (b) The two switching astroids for the nanomagnets in sublattices $L_a$ (solid outline) and $L_b$ (dashed outline), where $h_k$ is the coercive field strength along the hard axis, and $h_x$ and $h_y$ are the horizontal and vertical components of an external field.
    The astroid boundaries indicate the required magnetic field to switch a magnet in the corresponding sublattice.
    Astroid edges are coloured according to the magnet state promoted when crossing the edge.
    Similarly, the shaded areas indicate regions where selective switching occurs, promoting a single magnet colour from a single sublattice.
    In the centre of the astroids, the two bipolar clocks A and B are depicted at $+22\degree$ and $-22\degree$, respectively.
    }
    \label{fig:astroid-clocking}
\end{figure*}

Pinwheel ASI, depicted in \cref{fig:astroid-clocking}a, is composed of two interleaved square sublattices $L_a$ and $L_b$, whose nanomagnets are rotated $+45\degree$ and $-45\degree$, respectively.
In this work, we refer to the sublattice and magnetisation of the nanomagnets by their colour (blue or orange for $L_a$ and green or pink for $L_b$), as indicated by the coloured inset in \cref{fig:astroid-clocking}a.
Pinwheel ASI exhibits long-range ferromagnetic order, supporting large domains of coherent magnetisation.
We consider a $50\times50$ array with a background domain of leftwards magnetisation (blue-green).
In the centre of the array we initialise some magnetic structure with magnetisation in the opposite direction (orange-pink), which is considered the initial state of the ASI.

To drive the dynamics of the pinwheel ASI, we employ astroid clocking\citep{jensen2024clocked}.
\cref{fig:astroid-clocking}b shows the switching astroids for the nanomagnets in the two sublattices $L_a$ (solid outline) and $L_b$ (dashed outline).
A nanomagnet will switch (reverse magnetisation) if subject to a magnetic field that crosses the astroid boundary.
The shaded areas in the astroids indicate fields that selectively switch magnets from a single sublattice whose magnetisation is oppositely aligned to the field.

We define two clocking directions $A$ and $B$ along the $+22\degree$ and $-22\degree$ axes respectively.
$A$ and $B$ denote the positive (rightwards) clock field, while $a$ and $b$ refer to the negative (leftwards) clock fields.
$A$ and $a$ may selectively switch magnets from sublattice $L_a$, while $B$ and $b$ selectively switch magnets from $L_b$.
For example, applying the $A$ field will selectively switch blue magnets to the orange state.

Previous work has demonstrated that $AB$ clocking (pulsing the $A$ and $B$ fields in an alternating fashion) will result in step-wise growth of rightwards (orange-pink) domains.
Similarly, $ab$ clocking results growth of blue-green domains and consequently reversal of orange-pink domains.
For a domain contained within a larger outer domain, reversal is generally faster than growth\citep{jensen2024clocked}, when the field strengths are equal.
In general, domains can grow or shrink but do not move through the ASI, making this behaviour unsuitable for information transmission.

An orange-pink magnetic domain grows under positive fields $A$ and $B$, while shrinking under negative fields $a$ and $b$.
A glider moves while retaining its shape, and thus requires a balance of growing and shrinking.
Hence, we decouple the strength of the positive and negative clock fields, allowing for the possibility of a \emph{wonky} clock protocol.
We define $H^+$ as the strength of $A$ and $B$, and $H^-$ as the strength $a$ and $b$. 
We fix the clock protocol to $aAbB$, which can cause both growth and reversal within a clock cycle.
Glider discovery thus requires both an initial glider structure, and suitable field strengths.

We use an EA to efficiently search the large parameter space for regions which may support glider-like behaviour.
The EA is given the task of finding an initial ASI state along with suitable values for $H^+$ and $H^-$.
To evaluate a potential glider, we simulate its trajectory under the clock protocol using flatspin, the large-scale ASI simulator.
We define a fitness function which rewards glider-like behaviour, where a fitness of zero indicates a perfect score.
See Methods for further details.

\subsection*{The ``snake'' glider}
\begin{figure*}
    \centering
    \includegraphics[width=\textwidth]{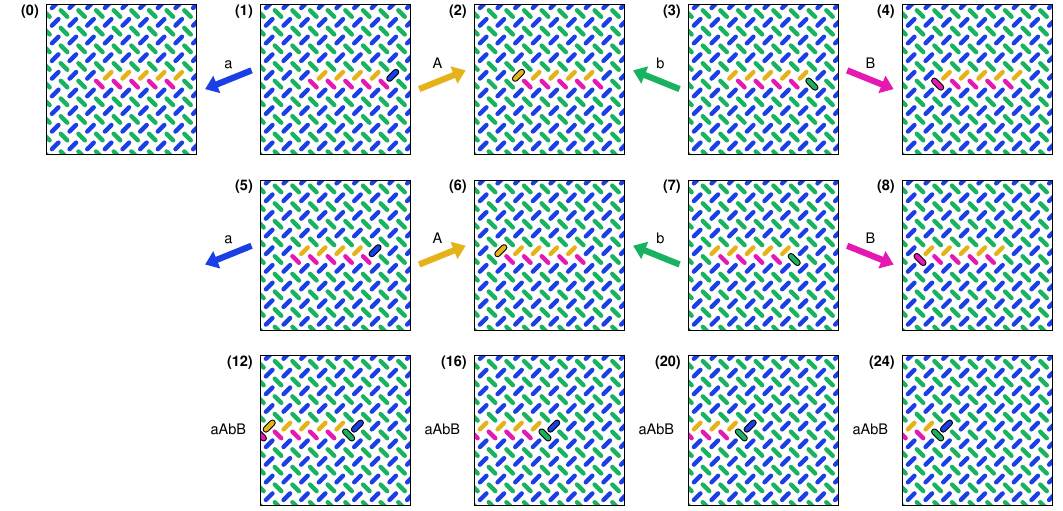}
    \caption{Snake glider in pinwheel ASI.
    Each snapshot shows a zoomed-in view of a $50\times50$ system, at different points during the $aAbB$ clock protocol.
    (0) shows the initial state of the snake, an elongated orange-pink domain in the centre of a blue-green background domain.
    (1-8) shows the state of the snake during $aAbB$ clocking, where magnets that change state between snapshots are highlighted by a solid black outline.
    The field arrow to the left of each snapshot indicates the clock field which precedes it.
    The snake moves leftwards through alternate shrinking ($a$ and $b$) and growing ($A$ and $B$) steps.
    The bottom row continues the series, showing only snapshots after a complete clock cycle (four clock fields are applied between each snapshot). Note that, despite the snake leaving the viewing window, it is still far from the edge of the ASI.
    }
    \label{fig:glider}
\end{figure*}

The EA discovered multiple interesting structures with glider-like properties.
Here we focus on the snake, shown in \cref{fig:glider}, the simplest glider discovered. 
The snake scores a perfect $0$ in the fitness function (\cref{eq:fitness}).
\Cref{fig:glider} (0) shows the initial state of the snake, consisting of a thin, elongated domain of orange and pink magnets.
\Cref{fig:glider} (1-8) show the time-evolution of the snake, where the snake is fully translated one step leftwards every four field applications (one full clock cycle). \Cref{fig:glider} (12, 16, 20, 24) show a continuation of the series, each a full clock cycle apart.
Further application of the clocking protocol will continue to translate the snake leftwards, until it reaches the edge of the ASI.
The behaviour of the snake resembles that of the 1D shift register constructed by Nomura et al. \cite{NomuraAPshiftReg2017}, though here it is able to function within a 2D system.
A video of the simulated snake glider is available in Supplementary Movie 1. 

The example in \cref{fig:glider} shows a leftwards moving snake, but surprisingly, if the snake is inverted with respect to the sublattices (pink magnets on the top, orange on the bottom), the snake then proceeds rightwards.
No changes to the clock protocol are needed for this reversal in direction, it is purely a function of the magnetic state.
Consequently, two snakes can inhabit the same ASI, experience the same clock protocol, but move in opposite directions, due only to their structure (see Supplementary information and Supplementary Movie 2).

We can understand the underlying mechanism of the snake glider by considering the effect of each individual field in the clock protocol. 
Let us refer to the ``pointy'' edge of the snake as its head, and the ``forked'' edge as its tail.
The snakes in \cref{fig:glider} are facing left (head on the left, tail on the right).
The $a$ and $b$ fields shrink the tail on the $L_a$ and $L_b$ sublattices respectively, while the $A$ and $B$ grow the head on the $L_a$ and $L_b$ sublattices, respectively.
In \cref{fig:glider} (1), the $a$ field switches the orange magnet at the tail, and ``unlocks'' the neighbouring pink magnet which is in turn switched by the $b$ field in \cref{fig:glider} (3).
A similar unlocking occurs for growth at the head.
Due to this unlocking mechanism, the clock fields can in fact be applied in any order, provided each is applied exactly once within a clock cycle.
Furthermore, the length of the snake can be altered by applying only the positive fields $A$ and $B$, causing the head to grow while the tail remains fixed.
Likewise, repeated cycling of the $a$ and $b$ fields causes the tail to shrink while the head remains fixed.
The snake can then be driven by the full clock protocol again and will retain its glider properties at its new size, provided it is not too small.
The smallest functional snake is of length two (two magnets on each sublattice).
There is no upper-bound on the length of the snake, provided it does not touch the edge of the ASI.

For verification of the snake glider, we reproduced the behaviour in the micromagnetic simulator MuMax3\cite{mumax3}.
These results are shown in the Supplementary information and corroborate the results of flatspin.
A video of the MuMax3 simulation is available in Supplementary Movie 3.

\subsection*{Experimental demonstration}
\begin{figure*}
    \includegraphics[width=\textwidth]{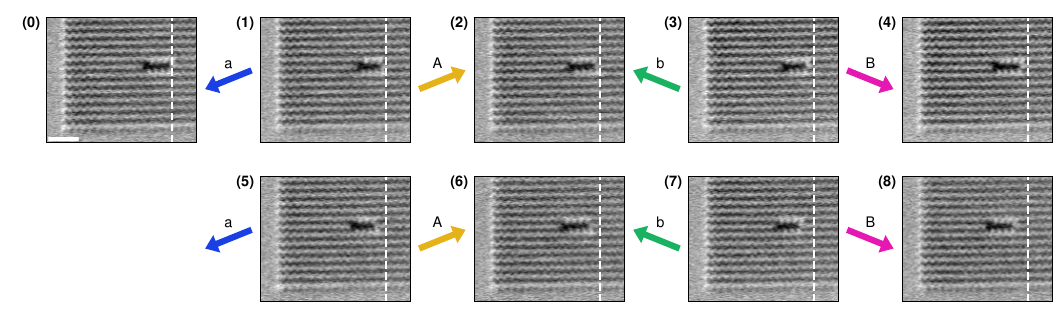}
    \caption{Experimental realisation of the snake glider in pinwheel ASI.
    Each MFM-micrograph shows a part of a \numproduct{100x100} pinwheel ASI, after application of a clock field in the $aAbB$ clock protocol.
    All MFM-micrographs are taken at 0 field.
    (0) shows the initial state of the snake, created by field-assisted writing using an MFM tip followed by a series of applied fields.
    The tail position of the initial snake is highlighted by the white dashed line.
    Scale bar: \qty{1}{\um}.
    (1-8) shows the movement of the snake during $aAbB$ clocking.
    }
    \label{fig:mfm-snake}
\end{figure*}

Next, we demonstrate snake movement experimentally (see Methods).
\cref{fig:mfm-snake} shows magnetic force microscopy (MFM) micrographs of the snake after each applied clock field.
The snake is visible in black, against the striped background of the polarised outer domain (blue-green in \cref{fig:glider}).
In the MFM-micrographs, the starting position of the tail is marked with a dashed white line.
\Cref{fig:mfm-snake} (0) shows the initial state of the snake.
In \cref{fig:mfm-snake} (1-8) we move the snake for two clock cycles, with behaviour identical to our simulations.
Comparing MFM-micrographs (0) and (8), the snake has moved leftwards by two lattice spacings, corresponding to \qty{\sim495}{\nm}.
A video of the full experimental series is available in Supplementary Movie 4.

The experimental realisation shows that snake movement is robust to inevitable variations in coercive fields, arising from fabrication defects and thermal noise. 
The glider behaviour may, however, break down when encountering nanomagnets with particularly large or small coercive fields (see Supplementary information).
Nevertheless, the results show that the snake can be used as a reliable information carrier.

\subsection*{Analysis}
While the EA discovered the snake at $H^+ = \qty{45.0}{\milli\tesla}$ and $H^- = \qty{32.9}{\milli\tesla}$, 
we find the snake to function reliably in flatspin simulation when $\qty{42.0}{\milli\tesla} \leq H^+ \leq \qty{45.5}{\milli\tesla}$ and $\qty{30.0}{\milli\tesla} \leq H^- \leq \qty{38.0}{\milli\tesla}$.
Notice these ranges enforce that $H^-$ is strictly less than $H^+$.
Applying a negative clock field with strength equal to the positive clock field switches more than a single tail nanomagnet, causing the snake state to disappear (see Supplementary information).
Intuitively, this can be understood by considering the two domains being grown.
The positive fields grow the smaller, inner domain (the snake), whereas the negative fields cause the larger, outer domain to collapse inwards.
The larger domain grows faster due to the morphology of the domain boundary, i.e., it envelopes the inner domain\citep{jensen2024clocked}.
It follows then, that the field for growing the larger domain ($H^-$) should be lesser in order to counterbalance the systems inherent bias towards reversal.

\begin{figure*}
    \includegraphics[width=\textwidth]{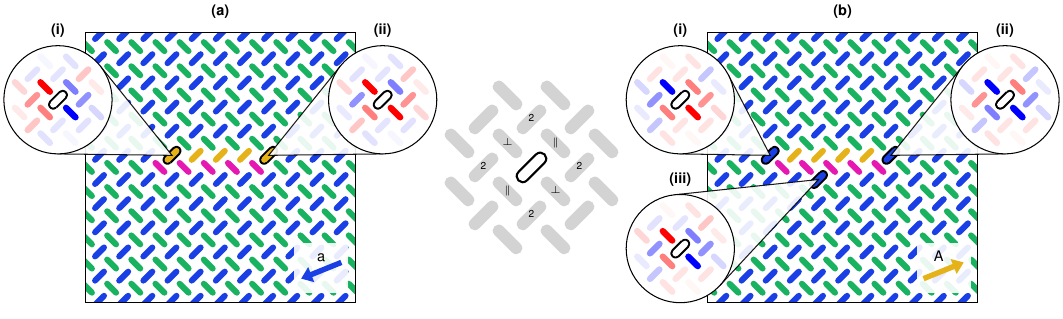}
    \caption{Neighbour influence during (a) shrinking and (b) growth.
    The five insets illustrate how the highlighted magnet is influenced by its neighbours through their dipolar fields.
    Stronger red signifies that the magnet is biasing the highlighted magnet towards switching, and stronger blue signifies that the magnet is biasing the highlighted magnet away from switching.
    The centre legend shows the parallel ($\parallel$) and perpendicular ($\perp$) nearest neighbours, as well as the second nearest neighbours ($2$).}
    \label{fig:snake-spotlight}
\end{figure*}

To further understand the mechanism of the snake's motion, we analyse the effect of each neighbour under an applied clock field.
In \cref{fig:snake-spotlight}, the five insets illustrate how the highlighted magnet is influenced by its neighbours through their dipolar fields.
Neighbour magnets with a stronger red colour are destabilizing, and effectively push the centre (highlighted) magnet towards the outside of the switching astroid.
A stronger blue colour indicates stabilizing interactions, with an effective pull towards the inside of switching astroid.
The centre legend of \cref{fig:snake-spotlight} depicts the local neighbourhood of a magnet.
We denote the four nearest neighbours of a magnet as $NN$, and the the four next-nearest neighbours as $NNN$.
Additionally, we distinguish between $NN_\parallel$ and $NN_\perp$, corresponding to the $NN$s that lie parallel or perpendicular to the magnet's principal (easy) axis.

The positive fields $A$ and $B$ are stronger than the negative fields $a$ and $b$.
Growth occurs during the positive fields $A$ and $B$, which we find are sufficiently strong to switch the entire sublattice, in the absence of dipolar fields.
Hence, during growth, dipolar interactions have a net stabilizing effect, preventing most magnets from switching.
In contrast, the negative fields $a$ and $b$ responsible for shrinking, are not strong enough to flip any magnets alone.
In this case, switching is mediated by dipolar interactions that have a net destabilizing effect, causing some magnets to switch.

\Cref{fig:snake-spotlight}a depicts the snake under the $a$ field, which effectively shrinks the tail of the snake by switching the rightmost orange magnet.
The insets for the two highlighted magnets at the head and tail reveal why only the tail magnet switches.
For the head magnet (\cref{fig:snake-spotlight}ai), the influence from the four $NN$ magnets effectively cancel out.
(The situation is similar for the orange magnets at the top of the snake.)
Even though three of the four $NNN$ magnets are in a weakly stabilizing configuration, the stability can be attributed primarily to the weakness of the $a$ field, where no switching occurs in the absence of strong destabilizing interactions.
However, for the tail magnet (\cref{fig:snake-spotlight}aii), there are indeed strong destabilizing interactions from its two $NN_\perp$ magnets (bright red).

\Cref{fig:snake-spotlight}b depicts the situation during growth of the head of the snake, where only the left-most highlighted blue magnet will flip under the $A$ field.
Now there are three relevant cases that could switch, namely the blue magnets at the head, below and at the tail of the snake.
As can be seen in the inset \cref{fig:snake-spotlight}bi, the head magnet loses stability when the $NN_\perp$ interactions cancel out, and is destabilized even further by its $NN_\parallel$ neighbours.
Meanwhile, the tail magnet (\cref{fig:snake-spotlight}bii) is strongly stabilized by its $NN_\perp$ neighbours.
Interestingly, the magnets below the snake also lose stability from $NN$ interactions that cancel out (\cref{fig:snake-spotlight}biii), but are weakly stabilized by three of the four $NNN$ magnets.

For the other sublattice ($L_b$) and corresponding fields ($B$ and $b$), the above analysis is identical (see Supplementary information).

Our analysis above shows that, under the applied clocking fields, the $NN_\perp$ magnets exert the strongest influence on switching (with strongest red or blue colour).
This may come as a surprise, since the dipolar fields from these neighbours are oriented perpendicular to the easy axis of the centre magnet.
However, due to the sloped edges of the switching astroid (\cref{fig:astroid-clocking}b), a perpendicular field can help push a magnet outside the astroid edge.
Hence, the clocked dynamics of the snake can not be understood in terms of dipolar energy alone (which only considers the parallel component of the dipolar field).

The apparent leftwards movement is a result of strong interactions from the $NN_\perp$ magnets of the tail magnet, and somewhat weaker interactions from the $NN_\parallel$ magnets of the head magnet.
In other words, the direction of movement is a result of the ``pointy'' shape of the snake's head and the ``forked'' shape of its tail.
If the snake is inverted with respect to the sublattices, the pointy head will be on the right side.
As a result, the mirrored snake will move towards the right.

\section*{Discussion}

To the best of our knowledge, the snake glider is the first example of the precise and controlled translation of a domain in ASI.
Previous work has shown how domains in ASI can be grown and shrunk in a reliable fashion\citep{jensen2024clocked}.
Gliders add a fundamentally new phenomenon to leverage in the control of magnetic structures.

The discovery of the snake glider provides a method for precisely controlling the magnetic state of an ASI.
This approach enables new directions in controlling other magnetic phenomena, such as creating channels for spin-waves in reconfigurable magnonic crystals\cite{li_writable_2022} or guiding the movement of magnetic nanoparticles\cite{gunnarsson_programmable_2005}.  

From a computational standpoint, the snake provides a reliable means of information transfer and memory.
Input can be encoded by nucleating snakes, which will propagate throughout the ASI, serving as a type of short-term memory.
ASI has previously been shown to be a capable substrate for transforming and manipulating data. 
With the addition of the glider as information carrier, the key fundamental criteria for computation\cite{LANGTON91} are fulfilled: the transmission, storage and modification of information.

Equipped with memory, ASI reservoirs can tackle temporal computing tasks directly, without the need for costly peripheral aids.
As a result, ASI reservoirs can be made vastly more energy efficient, avoiding the bottleneck imposed by the need for frequent reading and writing of reservoir state.

Going beyond, the snake could serve as the information carrier between larger complex systems, just as the glider in Game of Life based computers.
Snake-like structures could be employed to support communication between neural units in magnetic neuromorphic computing devices.
Integrating the modification, storage and transmission of data onto the same substrate, eliminates the need to translate information between mediums, unlocking significant efficiency gains.
With the discovery of the magnetic glider, ASI provides all the fundamentals for an all-magnetic computing chip.

\section*{Methods}

\subsection*{Evolutionary Algorithm}

The EA begins with a random population of individuals (solutions).
As the algorithm progresses, the individuals are assessed through simulation. The better performing individuals are retained, with a chance of variation (mutation or recombination).
Thus, through a process akin to natural evolution, the population undergoes iterative improvement.
The EA was run with a population size of $100$ for $100$ generations.

Individuals in the EA specify two field strengths $H^+$ and $H^-$, and an initial state. 
The EA has control of the initial state of a roughly-square patch of $128$ magnets at the centre of an ASI.
The initial state was represented as a one dimensional list of real values, with each element corresponding to one of the magnets in the centre square region of the ASI.
In close analogy to CA theory, we define the orange or pink spin state as \emph{on}, and the blue or green as \emph{off}.
A value of greater than $0.5$ in the initial state list, causes the corresponding magnet to be initialised in the \emph{on} state (polarised rightwards), all other magnets begin in the \emph{off} state.

To evaluate an individual we simulate its trajectory under the $aAbB$ clock protocol using flatspin\cite{JENSENflatspinPBR2022}.
The trajectory is the time-evolution of the ASI state, assembled by sequentially applying the fields of the protocol and recording the resulting spin state of the full ensemble.
We define the following fitness function on the resulting trajectory, to be \emph{minimised} by the EA: 
\begin{equation}
    \left(\sum_{t=t_0}^{T} |a_t - a_{t_0}|\right) + k \times (N - n_u)
    \label{eq:fitness},
\end{equation}
where $a_t$ is the number of magnets in the \emph{on} state at time $t$, $k$ is the penalty factor, $n_u$ is the number of unique ASI states in the period $t \in [t_0,T]$ and $N$ is the length of the trajectory ($T-t_0$).

In minimising this function, the EA rewards individuals which maintain a close to constant number of magnets in the \emph{on} state.
Additionally we employ a penalty when the number of unique states in the trajectory is lower than the total length of the trajectory, e.g, the trajectory settles into looping or stationary behaviour.
This penalty deters the EA from achieving a constant number of active magnets by utilising undesirable behaviour, such as frozen or oscillating dynamics.
We use a penalty factor of $k=5100$, equal to the number of nanomagnets in our simulated pinwheel ASI.  
A fitness score of $0$ indicates perfect glider behaviour has been attained.

For mutation, the algorithm selects whether to mutate $H^+$, $H^-$ or the initial state with even probability. Field strengths are mutated using Gaussian mutation with standard deviation $0.025$,  with the resulting field bounded to the range $[0.025, 0.05]$. For mutation of the initial state list, one element in the list is selected at random and mutated using Gaussian mutation with standard deviation $\frac{2}{3}$, with the result bounded to the unit interval.

For the recombination of two individuals, point crossover is applied to the initial state list to produce two offspring. 
Each offspring copies the field strengths of a different parent.  

A mutation rate of $0.2$ and a crossover rate $0.3$ is used. In each generation 20\% of the population is chosen for mutation and 30\% is chosen for crossover. 
The mutants and offspring are then evaluated for fitness. Selection considers the offspring, mutants and unaltered population and retains the $100$ individuals with best fitness. 

\subsection*{flatspin simulations}
Simulations are carried out using the flatspin point-dipole ASI simulator \citep{JENSENflatspinPBR2022}. In each simulation, the clock protocol $aAbB$ was sequentially applied 20 times to a $50 \times 50$ pinwheel ASI consisting of $5100$ nanomagnets.
The switching astroid of the simulated nanomagnets correspond to that of stadium shaped nanomagnets of dimension \qtyproduct{220 x 80 x 10}{\nano\meter}.

The flatspin simulations, used as part of the EA and in the subsequent field strength exploration, were run with the following parameters:
$\alpha=0.0025$, $b=0.4040$, $c=1.0$, $\beta=1.7331$, $\gamma=3.5195$, $h_k=0.1303$, $\text{neighbor distance}=10$, $\text{lattice spacing}=1$.

The values of $\alpha, b, c, \beta, \gamma$ and $h_k$ are taken from the astroid database provided with flatspin, for stadium magnets of size \qtyproduct{220 x 80 x 10}{\nano\metre}, with a saturation magnetisation value of \SI{860}{\kilo\ampere\per\meter}. 

Applying fields to an ASI often causes nucleation at the edges which grows inwards, due to the edge magnets having fewer close neighbours stabilising them.
However we wish the nucleation to be governed by the evolved initial state rather than the edges.
To this end, we add two layers (one of each sublattice) of \textit{buffer} magnets around the edge of the structure which are made harder to switch by increasing $h_k$ by a factor $10$.

\subsection*{Sample fabrication}

For experimental demonstration, we fabricated a \numproduct{100x100} pinwheel ASI consisting of \qtyproduct{220x80x10}{\nm} stadium-shaped Permalloy nanomagnets, with a lattice spacing of \qty{247.5}{\nm}.
Two layers of buffer edge-magnets were realised using \qtyproduct{220x70x10}{\nm} stadium-shaped nanomagnets, where the altered aspect ratio results in a higher switching field.

The pinwheel ASI with buffer edge-magnets was fabricated using an electron beam lithography lift-off process.
A 1:2 CSAR 62:anisole electron resist mixture was spin coated onto a Si substrate at 3250 rpm, resulting in a \qty{90}{\nm} thick electron resist layer. 
We then soft-baked the sample at \qty{150}{\degreeCelsius} for \qty{1}{\minute}, before exposing the desired pinwheel ASI using an Elionix ELS-G100 EBL system. 
Following exposure, the resist layer is developed in AR600-546 for \qty{40}{\second}. 
We then deposited a \qty{10}{\nm} layer of Permalloy (Ni$_{0.81}$Fe$_{0.19}$) using a K.J. Lesker E-beam evaporator.
Finally, ultrasound-assisted lift-off was performed in AR600-71.

\subsection*{Experimental demonstration}

We experimentally demonstrate movement of the snake with the $aAbB$ protocol by applying clock fields using an in-plane quadrupole vector magnet, where each magnetic field is applied for a couple of seconds.
The magnetic state of the pinwheel ASI is imaged using magnetic force microscopy (MFM) after each applied field.

We first polarise the pinwheel ASI with a \qty{60}{\milli\tesla} field pulse along \ang{180}. 
The snake state is initialised by writing a single \qty{\approx1}{\um} line, placed between sublattice $L_a$ and $L_b$, with the MFM-tip in the presence of a \qty{10}{\milli\tesla} bias field along \ang{0}. 
The \qty{10}{\milli\tesla} bias field is not strong enough to switch any nanomagnets on its own, but facilitates consistent and reliable magnetic writing with the MFM-tip. 
Writing was performed with the MFM-tip in contact with the sample, at a scan speed of \qty{55}{\um\per\s}.
We then apply a series of in-plane fields at the relevant clock field angles but with slightly varying field strengths to overcome some particularly hard to switch magnets and to optimise the field strengths used for clocking. 

We move the snake with the $aAbB$ field protocol with $H^+ = \qty{21.5}{\milli\tesla}$ and $H^- = \qty{18}{\milli\tesla}$ for two whole clock cycles. 
As expected, the experimental fields are weaker than those used in simulation, likely due to edge-roughness, oxidation and the thickness-dependence in the saturation magnetisation\citep{zhang2019}, none of which are taken into account in our simulations.
In the third clock cycle, some easily switched magnets also switch close to the tail of the snake, and the movement breaks down. 
We attribute the variation of switching fields of some of the magnets in the array to fabrication defects.

Both writing the initial state and imaging were done using commercial MFM-probes (NanoWorld POINTPROBE MFMR). 
MFM-imaging was done at remanence, with \qty{60}{\nm} lift height and \qty{55}{\um\per\s} scan speed. 
All experiments are performed at room temperature.

\section*{Data Availability}
The MFM data and MuMax3 input file used in this study have been deposited in the Zenodo database at \url{https://doi.org/10.5281/zenodo.14170198}.

\section*{Code Availability}
Numerical simulations were performed using the open-source simulators flatspin (\url{https://flatspin.gitlab.io/}) and MuMax3 (\url{https://mumax.github.io/}).


\section*{Acknowledgments}
This work was funded in part by the Norwegian Research Council TEKNOKONVERGENS project SPrINTER (Grant no. 331821), and in part by the EU FET-Open RIA project SpinENGINE (Grant no. 861618).
Simulations were executed on the NTNU EPIC compute cluster\cite{Epic2019}. The Research Council of Norway is acknowledged for the support to the Norwegian Micro- and Nano-Fabrication Facility, NorFab, project number 295864.

\section*{Author contributions}
AP did the initial discovery and designed the evolutionary experiment, with contributions from GT.
AP and JHJ designed the simulation study and performed the analysis.
IB fabricated the samples and did the experimental demonstration.
EF and GT provided feedback and suggestions.
AP, JHJ and IB wrote the manuscript with input from all authors.

\section*{Competing interests}
The authors declare no competing interests.

\end{document}